\documentclass[a4paper,conference]{IEEEtran}

\usepackage{amsmath}
\usepackage{array}
\usepackage{amssymb}
\usepackage{color, colortbl}
\usepackage{ulem}
\usepackage{enumitem}

\definecolor{LightCyan}{rgb}{0.88,1,1}
\definecolor{Gray}{gray}{0.9}

\usepackage{subfig}
\usepackage[pdftex]{graphicx}
\DeclareGraphicsExtensions{.pdf,.jpeg,.png}

\begin{document}
%
\title{SAGE: Sequential Attribute Generator for Analyzing Glioblastomas Using Limited Dataset}




%
\author{\IEEEauthorblockN{Padmaja Jonnalagedda\IEEEauthorrefmark{1}\IEEEauthorrefmark{4},
Brent Weinberg, MD, PhD\IEEEauthorrefmark{2},
Jason Allen, MD, PhD\IEEEauthorrefmark{2},
Taejin L. Min, MD\IEEEauthorrefmark{2},\\
Shiv Bhanu, MD\IEEEauthorrefmark{3} and
Bir Bhanu, PhD\IEEEauthorrefmark{1}}
\IEEEauthorblockA{\IEEEauthorrefmark{1}Dept. of Electrical and Computer Engineering, University of California at Riverside, Riverside, California\\}
\IEEEauthorblockA{\IEEEauthorrefmark{2}Dept. of Radiology and Imaging Sciences, Emory University, Atlanta, Georgia\\}
\IEEEauthorblockA{\IEEEauthorrefmark{3}Dept. of Radiology, Riverside Community Hospital, Riverside, California\\}
\IEEEauthorblockA{\IEEEauthorrefmark{4}Email: sjonn002@ucr.edu\\}
}

\maketitle

\begin{abstract}
While deep learning approaches have shown remarkable performance in many imaging tasks, most of these methods rely on the availability of large quantities of data. Medical imaging data, however, are scarce and fragmented. Generative Adversarial Networks (GANs) have recently been very effective in handling such datasets by generating more data. If the datasets are very small, however, GANs cannot learn the data distribution properly, resulting in less diverse or low-quality results. One such limited dataset is that for the concurrent gain of 19/20 chromosomes (19/20 co-gain), a mutation with positive prognostic value in Glioblastomas (GBM). In this paper, imaging biomarkers are detected for the mutation to streamline the extensive and invasive prognosis pipeline. Since this mutation is relatively rare, i.e. small dataset, a novel generative framework – the Sequential Attribute GEnerator (SAGE), is proposed, that generates detailed tumor imaging features while learning from a limited dataset. Experiments show that not only does SAGE generate high quality tumors when compared to Progressively Growing GAN (PGGAN), Wasserstein GAN with Gradient Penalty (WGAN-GP) and Deep Convolutional-GAN (DC-GAN), but also captures the imaging biomarkers accurately.
\end{abstract}

\IEEEpeerreviewmaketitle

\section{Introduction}

\label{sec:intro}

With recent attempts towards Assistive AI and Computer-Aided Diagnosis in the medical world, a common problem encountered is the lack of curated data to train the networks. Generative Adversarial Networks (GANs) have been shown to learn the data distribution from available data and generate a variety of data, which can serve as additional data for training. However, GANs also need data to train on and the need for data to create more data poses a classic causality dilemma. In the proposed research, we demonstrate a technique to generate tumor images using limited dataset, by expanding the latent representation of the tumor features. Such an approach would prove significant in studying rare mutations such as 19/20 co-gain – an indicator of prognosis in brain tumors. 

\textbf{Clinical Background:} Glioblastoma multiforme (GBM) is the most common and aggressive form of malignant tumor, comprising of 54\% of all primary brain tumors \cite{aans}, reporting a 5-year survival rate of 5\% \cite{tamimi2017epidemiology}. A significant statistic is that the survival rate has not improved in the last three decades \cite{tamimi2017epidemiology}, which therefore, brings to attention the need for accurate evaluation of prognosis and efficacy of chemotherapy. Assessment of overall clinical outcomes typically requires a combination of clinical, molecular and multi-modal imaging data. This process is time consuming, invasive, cumbersome and overloads the clinical workforce. Some of the many factors that contribute to this are increasing incidence of GBM, high resolution imaging, paucity of resources for molecular testing, lack of follow-up, inconsistent data recording across modalities, etc. As an answer to this problem, researchers have started looking into optimizing the radiology pipeline by estimating prognostic markers in imaging, in an attempt to bypass the overhead caused by molecular data collection and analysis. Among the many prognostic markers \cite{hegi2005mgmt, hartmann2013long, geisenberger2015molecular}, the 19/20 co-gain is yet to be explored. In this paper, we show that there are discriminatory imaging biomarkers indicating mutation and that we can recreate them using SAGE. 

\begin{figure}[t]
   \begin{center}
   \includegraphics[width=0.98\linewidth]{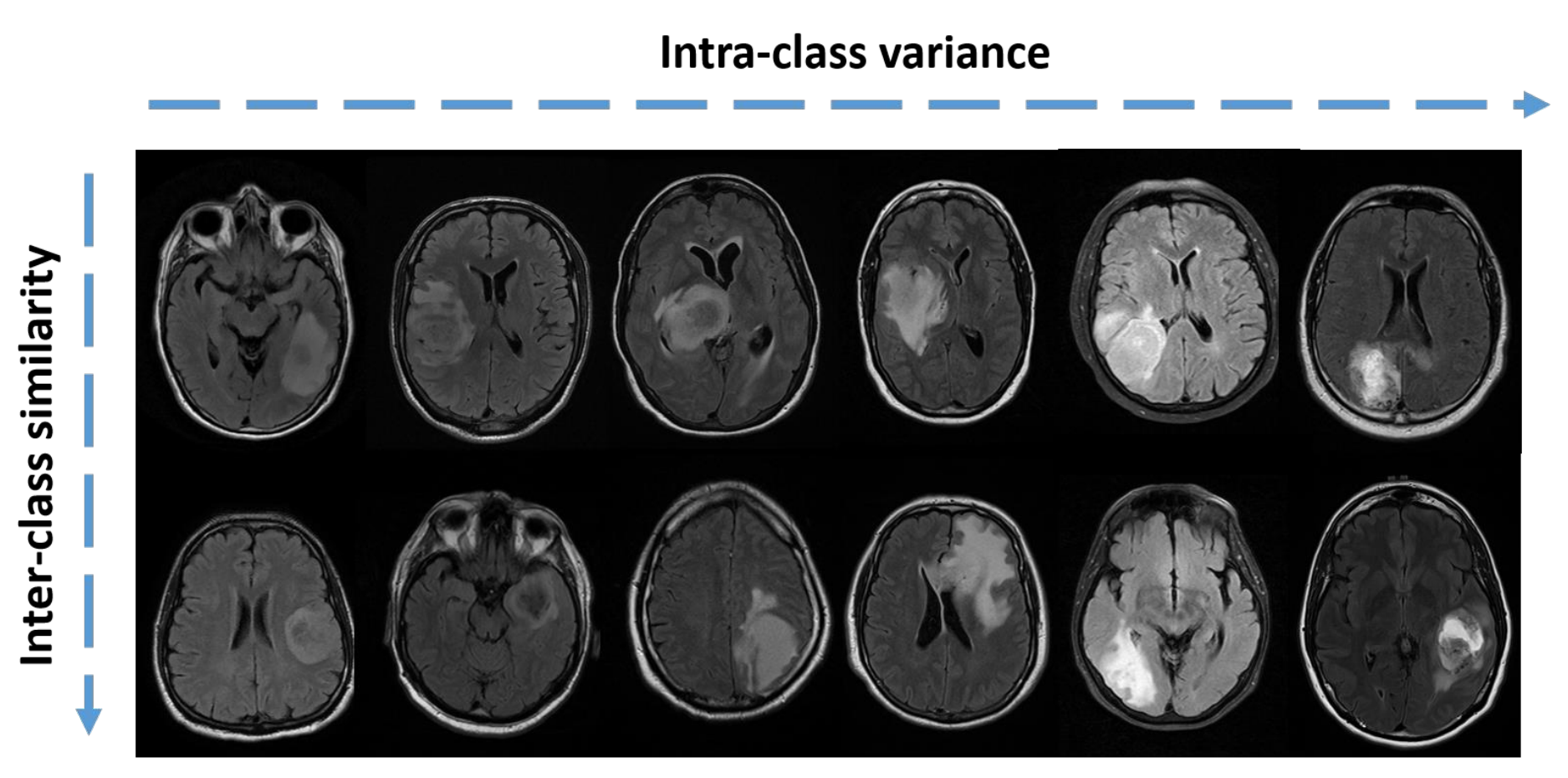}
   \end{center}
   \caption{Intra-class variability and inter-class similarity in 19/20 co-gain mutation. Images are all shown in FLAIR (fluid-attenuated inversion recovery) MR modality. Top: Mutation absent, Bottom: Mutation present}
   \label{fig:var}
\end{figure}

\begin{figure*}[t]
   \begin{center}
   \vspace{-5mm}
   \includegraphics[width=0.9\linewidth]{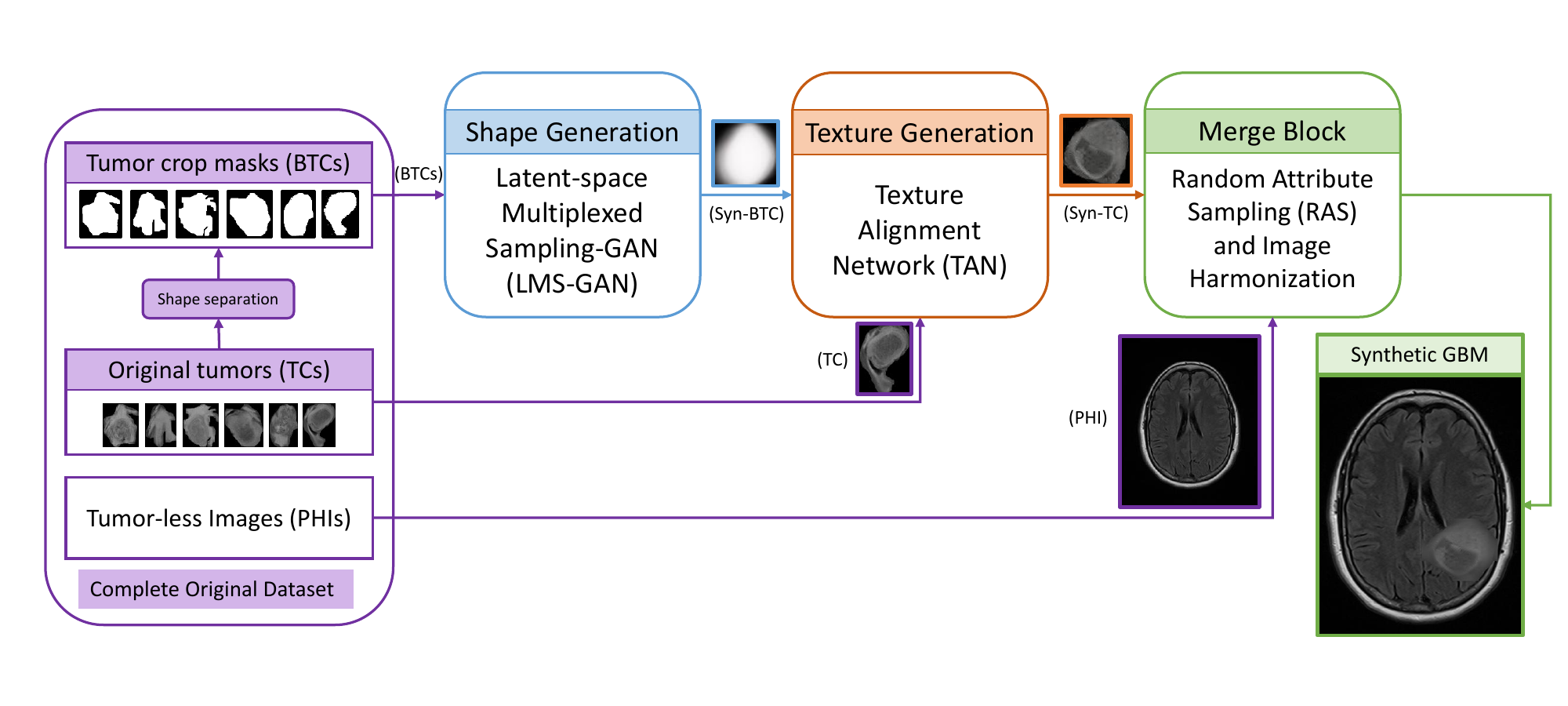}
   \end{center}
   \vspace{-5mm}
   \caption{Overall framework of the Sequential Attribute GEnerator (SAGE). BTC, TC and PHIs are the isolated features of real data which are re-sampled in the Shape Generation, Tumor Generation and Merge (Location generation) blocks, respectively.}
   \label{fig:sage}
   \vspace{-3mm}
\end{figure*}

\textbf{Motivation for SAGE:} One of the major challenges in the medical analysis is lack of data. Since 19/20 co-gain is a rare mutation, only a limited dataset is available. Furthermore, it also has a high inter-class similarity and intra-class diversity (Fig~\ref{fig:var}), which makes the visual assessment of biomarkers very unapparent to the naked eye. Any analysis using GANs will be successful only if they are capable of generating synthetic images that capture the nuance and diversity of features in the mutation. This implies the existence of a data distribution that GANs can learn from. Additionally, since 19/20 co-gain has never been analyzed before for visual manifestation in MR Imaging (MRI), we begin by demonstrating that there is, indeed, a presence of consistent imaging biomarkers that have a positive correlation with presence or absence of 19/20 co-gain. Following this, we will proceed to learn and recreate these biomarkers using SAGE. The motivation behind developing SAGE as opposed to using standard PGGAN, WGAN or DC-GAN is that SAGE is designed to increase the apparent size of the dataset. So to mitigate the issue of lack of data samples to train GANs, SAGE increases the size of dataset in latent space by deconstructing features and offers a wider feature search space, and thereby a better estimate of data distributions. We show that SAGE learns useful features from the dataset to generates good quality and diverse synthetic images. 

The rest of the paper is organized as follows: Section 2 contains related work and contributions. Section 3 details the technical approach. Section 4 contains the description of dataset and evaluation setup. Section 5 presents the results and Section 6 provides the conclusions.

\section{RELATED WORK AND CONTRIBUTIONS}
\label{sec:format}

\subsection{Related Work}
\label{ssec:related}

Deep learning has recently achieved remarkable success in various aspects of automated workflow involving MR Image Analysis such as diagnosis \cite{litjens2016deep}, grading \cite{citak2018machine}, segmentation \cite{pereira2016brain} and other clinical outcomes \cite{chang2018deep}. Generative Adversarial Networks (GANs) have recently gained attention in the medical field due to their efficacy in modeling and recreating data distributions to tackle common issues with medical datasets \cite{han2018gan, livergan1, livergan2}, the primary goal being data augmentation. For MRI, GANs have been used for data augmentation \cite{han2018gan}, segmentation \cite{zhang2018ms}, data anonymization \cite{shin2018medical}, etc. Most of these approaches use GANs to generate the whole image at once. However, this approach does not necessarily generate the most detailed images. Since we are analyzing unapparent visual manifestations, it is critical that the tumor features be generated with sufficient detail. For High- and Low-Grade Gliomas, there has been some research in generating images for data augmentation \cite{han2018gan, han2019learning}. Some of the research has attempted to use manifold learning \cite{khayatkhoei2018disconnected}, attention-based learning \cite{zhang2018self}, Variational Autoencoders \cite{higgins2017beta}, etc. for improving learning quality. Real-world medical data, however, reside in a higher dimension with many complex attributes operating together. Adding to this complexity, medical data are often fragmented and scarce. With SAGE, we establish the impact of separating features and latent space re-sampling to increase the apparent sample space of the dataset to match the complexity and dimensionality of the data. This results in significantly better-quality images.

\subsection{Contributions}
\label{ssec:contributions}

\begin{itemize}[leftmargin=*]
    \setlength \itemsep{0.2em}
\item A novel approach that can generate diverse synthetic images from very limited datasets using feature recasting,
\item Feature disentanglement and sequential generation for high-resolution images and added control over generated tumor properties,
\item Quantitative analysis of efficacy of proposed method in learning and recreating visually unapparent data distribution compared to naive GANs
\end{itemize}

\section{TECHNICAL APPROACH}
\label{sec:approach}

\subsection{Overview of Approach}
\label{ssec:overview}

The Sequential Attribute GEnerator (SAGE) framework (Figure~\ref{fig:sage}) has three modules: (a) Shape Generation module, (b) Texture Generation module and (c) Merge module. Essentially, SAGE generates tumor crops using (a) and (b) and then merges the tumors with tumor-less brain slices in (c). The detailed explanation of each of these modules is given in Section ~\ref{ssec:framework}. From the dataset, we separate the slices of MR Images that contain tumor vs the ones that do not. Tumors are segmented manually from the slices that contain them to create Tumor Crops (TC). Developing deep learning approaches for brain tumor segmentation is an actively growing field. Some of the best performing networks \cite{myronenko20183d} will also have some errors. If a tumor is wrongly segmented, the error will be propagated through the other modules of SAGE, generating incorrect tumors. We, therefore, opted for manual segmentation to obtain TCs. The TCs are then subjected to simple binary thresholding to obtain the shape masks. These will be referred to as Binary Tumor Crops (BTCs), henceforth. The slices that do not contain tumors are used in the merge block input (Tumor-less Images in Figure 2) and are called Pseudo-healthy Images (PHI). When a tumor grows in the brain, it pushes through the healthy brain tissue. Since there is yet no mathematical model that emulates this impact, we have used PHIs. PHIs are a part of the brain that contained a tumor and have a residual impact. Therefore, they are the closest images we can use for merging with the tumors realistically. The properties of tumors and PHIs correspond to where the tumor can be located in the brain. BTCs are a representation of shape of tumors whereas TCs have the texture properties. So, we now separate and re-sample from three macro-features: shape, texture and location. Recasting and re-sampling of these features gives us an apparent increase in sample space since we now have three data distributions instead of one. Figure~\ref{fig:sage} outlines the overall SAGE approach. The following sections provide details on the SAGE workflow.

\begin{figure}[b!]
   \begin{center}
   \vspace{-2mm}
   \includegraphics[width=0.98\linewidth]{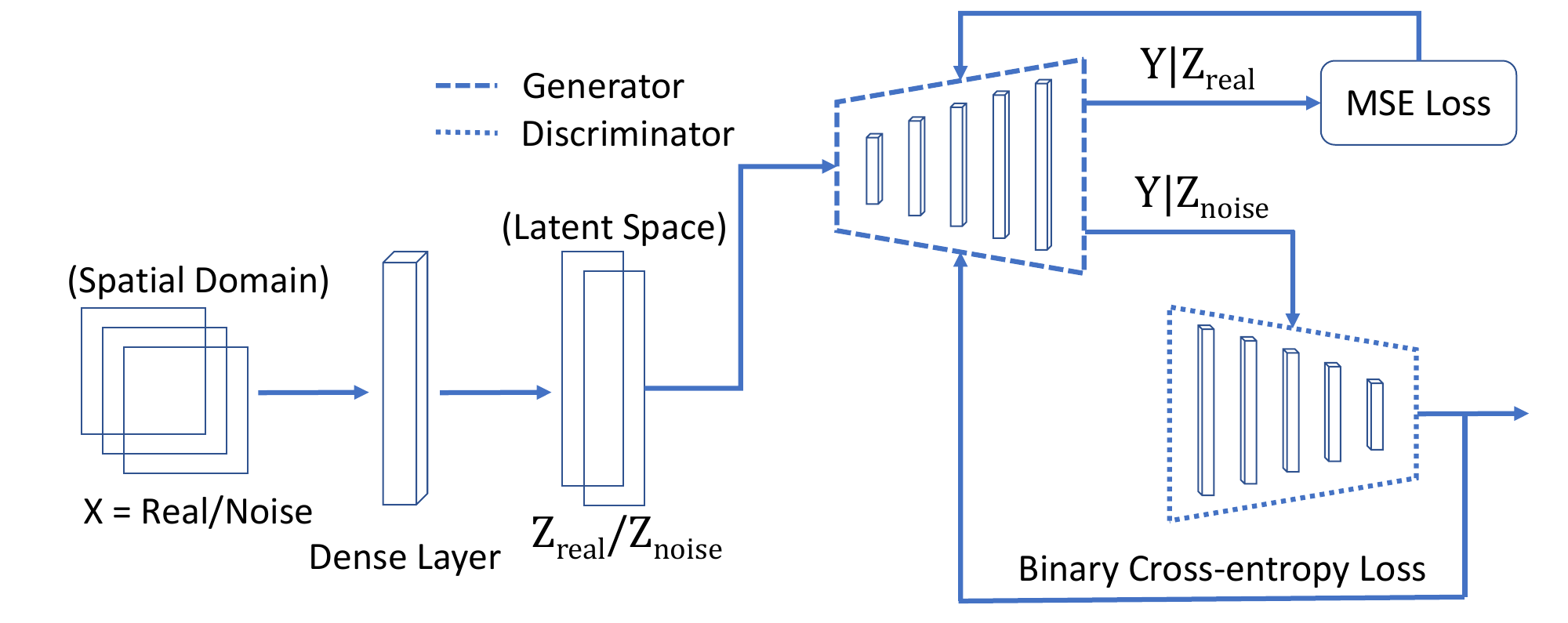}
   \end{center}
   \vspace{-3mm}
   \caption{Framework of Latent-space Multiplexed Sampling (LMS) GAN}
   \label{fig:lms-gan}
\end{figure} 

\subsection{Description of Framework}
\label{ssec:framework}

In this section, we will detail each of the modules of SAGE as described in Section~\ref{ssec:overview}. The shape generation module of SAGE consists of a Latent-space Multiplexed Sampling GAN (LMS-GAN). The texture generation module consists of a Texture Alignment Network (TAN).\medskip

\textbf{LMS-GAN:} The Latent-space Multiplexed Sampling GAN (LMS-GAN) is the shape generator module of SAGE. It is a densely connected GAN framework that multiplexes between latent space vectors of data and noise to generate a binary mask of the tumor. The input of LMS-GAN are the BTCs. LMS-GAN performs a non-linear mapping from the input image or noise to the latent space. From the latent space, the network randomly samples data points and gives it as an input to the generator, periodically switching the latent vector between noise ($Z_{noise}$) and real data ($Z_{real}$). By randomly sampling points from the latent space, it can generate as many variations of an input vector to the generator as needed. However, since the latent vector input to the generator is no longer a feature vector from a coherent image, the generator block is designed as a densely connected network and not a convolutional network. To train the generator, the loss function switches with the input data type. If the sampling is done from $Z_{real}$, the network is trained on Mean Squared Error (MSE) loss (Eq~\ref{eqn:L_real}) and if the samples are $Z_{noise}$, loss is Binary Cross-Entropy (BCE) loss (Eq~\ref{eqn:L_noise}). The overall loss function is shown in (Eq~\ref{eqn:sage-loss}). 

\begin{subequations}
    \begin{equation}
        \mathcal{L}_{X \sim p(real)}: \mathbb{E}_{x \sim p_{real}(x)}||x_{real}-y||_2
        \vspace{-4mm}
        \label{eqn:L_real}
    \end{equation} 
     
    \begin{equation}
        \vspace{-3mm}
        \mathcal{L}_{X \sim p(noise)}: \mathbb{E}_{x \sim p_{noise}(x)}\Big[\log \Big( 1-\mathbf{D}\big(\ \mathbf{G}(z_{noise}) \ \big) \Big)\Big]
        \label{eqn:L_noise}
    \end{equation} 
\end{subequations}

\begin{equation}
    \mathcal{L}_{LMS} = {\mathbb{I}_x} \times \mathcal{L}_{X \sim p(noise)} + ({1 - \mathbb{I}_x}) \times \mathcal{L}_{X \sim p(real)}
    \label{eqn:sage-loss}
\end{equation} 

where $\mathbb{I}_x$ is the indicator function which is 0 when $X$ is data and 1 when $X$ is noise. The noise is sampled from random normal distribution. $\mathbf{G}$ and $\mathbf{D}$ are Generator and Discriminator, respectively. $y$ is the generated image and $x_{real}$ is the real image. $Z_{real/noise}$ are the latent space representations of real and noisy inputs. This is depicted in Fig~\ref{fig:lms-gan}.\medskip

\textbf{Texture Alignment Network (TAN):} This module generates texture. TAN takes the Synthetic-BTCs (Syn-BTCs) generated by LMS-GAN and “assigns” a texture. TAN randomly samples a texture from the pool of TCs, learns its feature representation and aligns the distribution of Syn-TC with that of the sampled TC distribution. The output of this block is a synthetic tumor crop (Syn-TC). Some researchers have performed feature reassignment \cite{gatys2015neural, jing2019neural} in the past. In our approach, we use features from Conv3, 4, 5 blocks of VGG19 and train on perceptual loss (Eq~\ref{eqn:TAN-loss}). We only include the deeper layers because texture of an image is learnt in the deeper layers of a CNN.
\begin{equation}
    \mathcal{L}_{TAN}: \sum_{i} \omega_i \times ||\mathrm{g}_{i}(T) - \mathrm{g}_{i}(I)||_2 + ||\mathbb{F}_{i}(B) - \mathbb{F}_{i}(I)||_2
    \vspace{-1mm}
    \label{eqn:TAN-loss}
\end{equation} 

where: $\mathcal{L}_{TAN}$: overall loss function, $\omega_i$: weight of $i^{th}$ layer, T: sampled TC, B: input Syn-BTC, I: output Syn-TC, $\mathbb{F}_i$ output of $i^{th}$ layer and $\mathrm{g}_i$: Gram matrix of $i^{th}$ layer output.\medskip

\textbf{Merge Block:} When we merge the Syn-TCs with PHIs, we must ensure maximum diversity and uniformity to generate realistic images. For diversity, we randomly select location, relative size of tumor (with respect to whole image) and rotation angle for the Syn-TC. The values of these attributes are chosen from a pre-defined allowed range of values. This range is obtained from the mean and standard deviation of the original dataset attributes so we can choose attribute properties with a certain confidence. Once the tumor is merged, we apply edge-tapering Gaussian filters for smoothness. The overall framework of SAGE allows us to address the issues of image quality because of data scarcity commonly experienced by GAN approaches. In the following section, the results of using SAGE framework to generate GBM images are detailed.

\begin{figure*}[t]
  \begin{center}
  \includegraphics[width=0.9\linewidth]{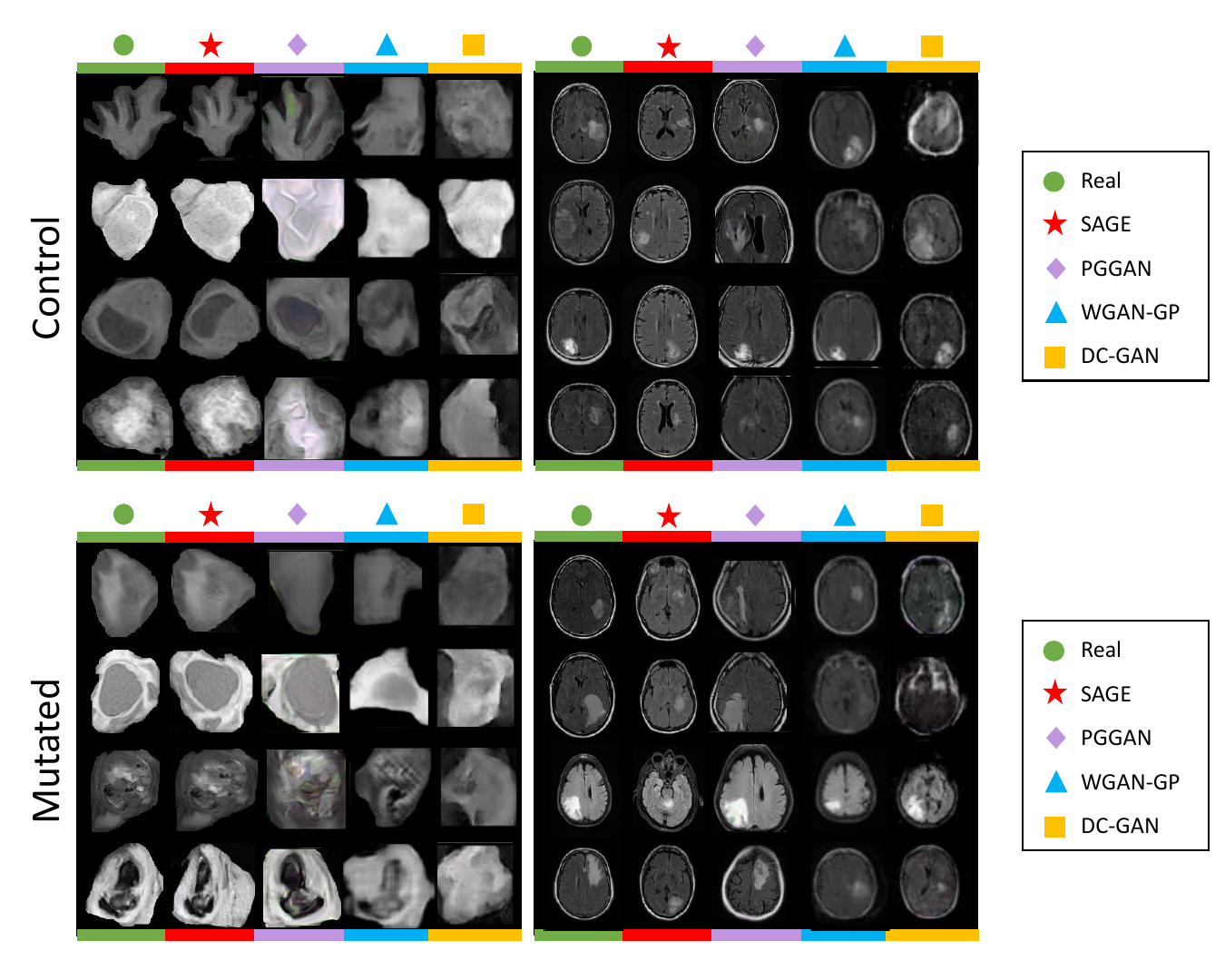}
  \end{center}
  \vspace{-5mm}
  \caption{Tumor crops and Whole Images generated by SAGE vs Real data (compared against PGGAN, WGAN-GP and DC-GAN)}
  \label{fig:sage-res}
\end{figure*}

\section{EXPERIMENTAL SETUP}
\label{sec:exp_setup}

\subsection{Description of Dataset}
\label{ssec:dataset}

We use a dataset containing FLAIR MR images for a cohort of 25 patients with known 19/20 co-gain status. The data is divided into two classes: control class that has 19/20 co-gain absent and the mutated class that has co-gain present. This dataset was provided by our collaborators at Emory University. The cohort is divided into 14 control and 11 mutated patients. Each patient has an average of only 9 FLAIR images with tumor, making it an average of ~110 images per class (without train-test split), this makes for an even smaller training set. Thus, this is a very limited dataset to train GANs. 

\subsection{Evaluation Protocol}
\label{ssec:evaluation}

\begin{figure*}[t]
   \begin{center}
   \includegraphics[width=\linewidth]{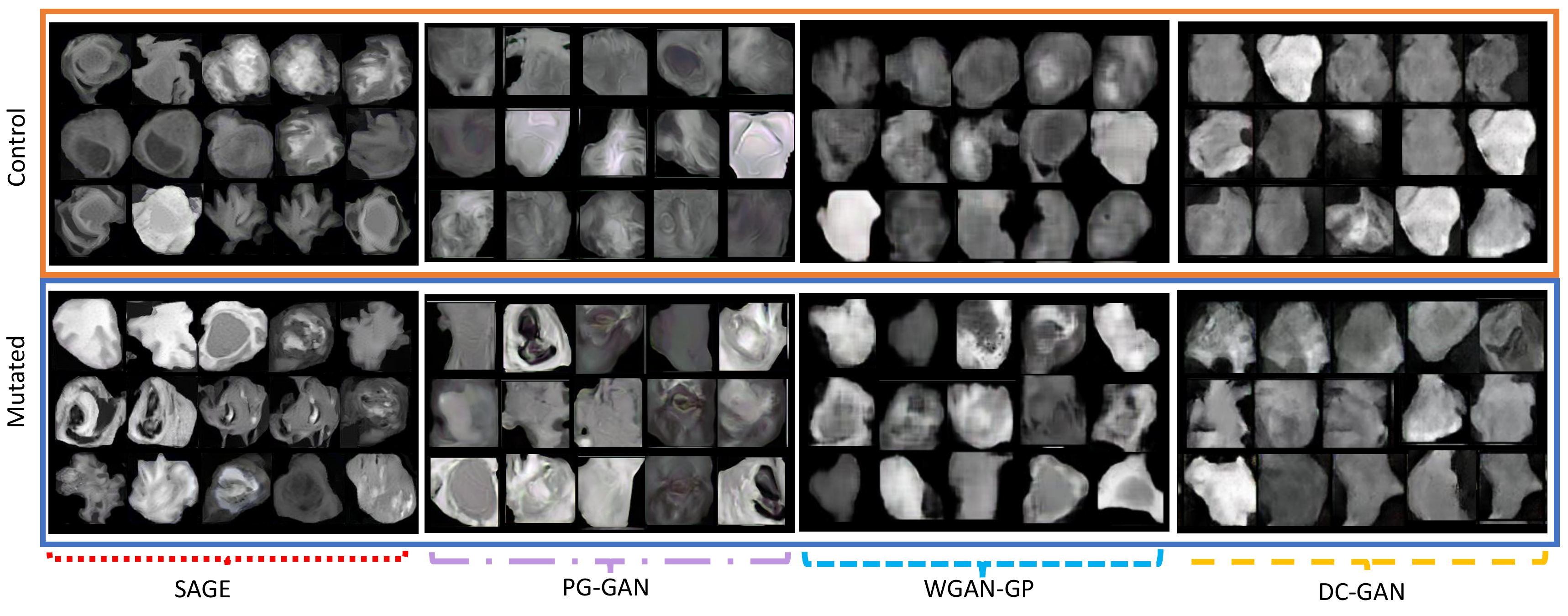}
   \end{center}
   \caption{Diversity of images generated by SAGE vs PGGAN vs WGAN-GP vs DC-GAN for both classes of tumors. The green (solid) box contains real samples, the red (dotted) box contains SAGE samples, purple (dot-dash) box: PGGAN samples, blue (dashes-medium): WGAN-GP and yellow (dashes-large): DC-GAN samples.}
   \label{fig:diversity-res}
\end{figure*}

\begin{figure*}[h]
   \begin{center}
   \includegraphics[width=0.75\linewidth]{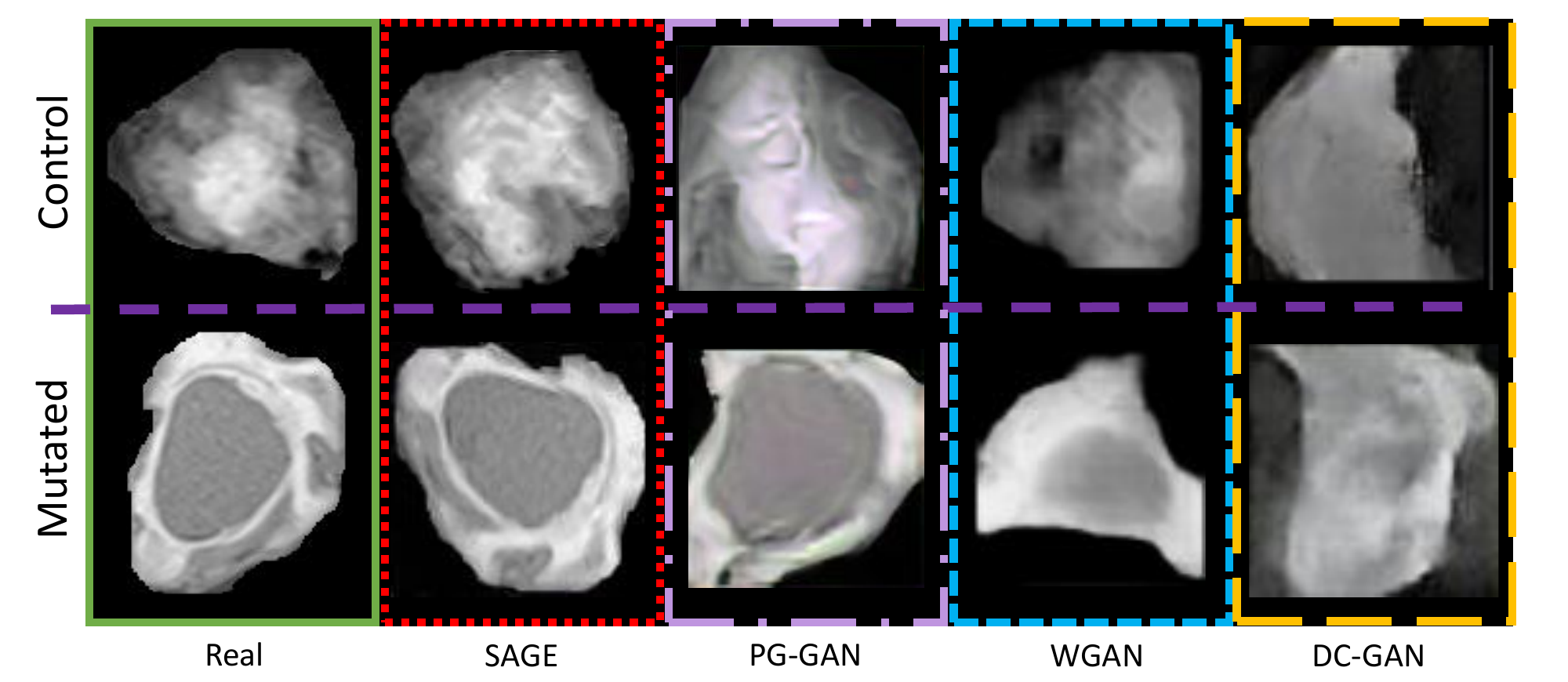}
   \end{center}
   \caption{Comparing quality of similar-looking generated images. The purple dashed line divides control and mutated samples. The green (solid) box contains real samples, the red (dotted) box: samples generated by SAGE, purple (dot-dash) box: PGGAN samples, blue (dashes-medium): WGAN-GP and yellow (dashes-large): DC-GAN samples.}
   \label{fig:zoomed-res}
\end{figure*}

\textit{1. Qualitative Evaluation:} For qualitative evaluation, we present three types of results, each of which demonstrates a specific strength of SAGE. One result evaluates how well SAGE can learn the inherent data distribution of the data and translate it into generating new tumors. This is shown by comparing SAGE generated images with real images as well as PGGAN, WGAN-GP and DC-GAN results. We chose four real tumor crops with very different visual properties for each class and picked corresponding generated images from all three GANs that most closely resembled these tumors. This provides evidence of how well SAGE can learn the properties of highly diverse and limited data. For the second type of results, we demonstrated how SAGE can generate a diverse range of synthetic tumors compared to other GANs. This is an evaluation of the robustness of SAGE towards mode-collapse and loss of detail in generated images. For the third type of results, we show how well SAGE can capture the quality and detail of tumors. Thus, we zoom in into an example set of images for each class and compare resolution and detail between real, SAGE, the other GANs.\medskip

\textit{2. Quantitative Evaluation:} Inception Score and Structural Similarity: We report Inception Score (IS) and Structural Similarity Index (SSIM) to demonstrate the quality of generated images. The Inception Score evaluates two properties: (a) Image quality and (b) Image diversity. To assess image quality, meaningful generated objects should have a conditional label distribution with low entropy \cite{salimans2016improved}. For ensuring diversity in generated images, the marginal probability should have high entropy \cite{salimans2016improved}. Inception score is a combination of these two constraints. It achieves this by calculating the KL Divergence between the conditional and marginal distributions. The average KL divergence value of all generated images is the score. The lowest value of IS is 1 and the highest value is the number of classes. The SSIM is a comprehensive measure of similarity between images. The value of SSIM lies between 0 and 1, where 0 means the images aren't similar at all. Ideally, we want our images to follow the same distribution as real data but also be different enough to ensure diversity. Therefore, we can identify the range of 0.35-0.75 as the preferred range of SSIM to evaluate our synthetic image quality.

\textit{3. Blind Test by Radiologists:} We conducted tests for assessing the quality of generated tumor crops via a blind prediction test with radiologists. We asked radiologists to manually distinguish between real and synthetic tumors. The evaluation was done with 3 radiologists. They were provided with control and mutated folders, each containing random number of real and synthetic images. These images were randomly chosen and resized to 100x100 for consistency. Furthermore, the names of these images were replaced by random numbers to hide any indicators that may point towards the nature of the image. For a completely blind assessment, the statistics of data split were not revealed to the radiologists. The control group had a total of 92 images: 45 real and 47 synthetic while the mutated group had 82 images: 35 real and 47 synthetic images. The radiologists were then asked to mark the images R/F (Real or Fake). These results are summarized in Table~\ref{tab:blind_test}. Here, accuracy (ACC) is the proportion of correct predictions made, False Positive Rate (FPR) is the proportion of synthetic images that were mistaken as real, False Negative Rate (FNR) is the proportion of real images mistaken as synthetic and Precision (PR) is the proportion of true positives over all positives. During our design of experiment for this test, there were two key questions posed by radiologists which we answered. These questions and answers are given below:

\begin{table}[t]
  \centering
  \caption{Inception score (IS) and Structural Similarity (SSIM) for generated images using SAGE, PGGAN, WGAN-GP and DC-GAN. The number of generated images per class is 500 for each GAN.}

  \begin{tabular}{c|cc}
  \bfseries Generative Model & \bfseries IS & \bfseries SSIM \\ \hline \hline
  \textbf{SAGE (ours)} & \textbf{1.71} & \textbf{0.68} \\
  PG-GAN & 1.55 & 0.66\\
  WGAN-GP & 1.35 & 0.57\\
  DC-GAN & 1.12 & 0.32 \\
  \hline
  \end{tabular}
  \label{tab:quality_metrics}%
\end{table}

Question 1: What is your hypothesis that you are trying to show by having us classify them? Do you predict that a human will or will not be able to tell the difference?

Answer 1: We want to have an expert evaluation of the quality of generated tumor images. If the generated images are of high quality, one will not be able to distinguish them from real images. Based on this, we can have a metric of the quality of generated images. This will also help us to improve the quality of generated images in the future.\medskip

Question 2: What is the reasoning for the tumor being so tightly cropped? For us, it is very unusual to look at an image that is only the segmented tumor. In this way, all the images seem very strange to us. It is difficult to reliably tell the difference because there is not much context.

Answer 2: The images are tightly cropped since only the tumor is (synthetically) generated. Cropped images are used for data augmentation for the training of deep learning algorithms (to distinguish mutated/control groups) assuming the detection/segmentation has been done perfectly. We aim to evaluate how well the tumor features are learnt when isolated. Yes, the context is very important for image interpretation and currently we take that into account in the classification step where we learn the discriminatory features.\medskip

\textit{4. Learning Discriminatory Features:} We focus our evaluation on our primary goal: to evaluate whether SAGE can consistently and precisely learn the desired feature distribution. To evaluate this, we use the generated images to classify real images. The rationale is: if SAGE-generated images can detect discriminatory features in a real test set, then SAGE has captured the desired data distribution unique to each class. We compare these results using synthetic images generated using the 3 other GANs to demonstrate the effectiveness of SAGE in learning from limited data. We report accuracy, sensitivity, specificity and dice score for a complete evaluation of performance. Here, accuracy is the proportion of total correct predictions, sensitivity is the true positive rate, specificity is the true negative rate and dice score is the harmonic mean of precision (proportion of true positives over all positives) and recall (sensitivity).

\begin{table}[t]
  \centering
  \caption{Blind test for radiologists to distinguish between real and synthetic tumor crops. The metrics shown are: ACC - Accuracy, FPR - False Positive Rate, TNR - True Negative Rate, PR - Precision. Row (non-shaded) are values for each radiologist and row (shaded) is the mean value across all radiologists.}
 
  \begin{tabular}{c|c|cccc}
  \bfseries Data class & \bfseries Radiologist & \bfseries ACC & \bfseries FPR & \bfseries FNR & \bfseries PR\\ \hline \hline
  & 1 & 0.59 & 0.42 & 0.38 & 0.55 \\
  \textit{Control} & 2 & 0.67 & 0.34 & 0.31 & 0.66 \\
  & 3 & 0.70 & 0.30 & 0.36 & 0.64 \\
  \hline
  \rowcolor{LightCyan}
  & Mean & 0.66 & \textbf{0.35} & 0.33 & 0.64 \\
  \hline \hline
  & 1 & 0.74 & 0.33 & 0.19 & 0.72 \\
  \textit{Mutated} & 2 & 0.82 & 0.19 & 0.18 & 0.87 \\
  & 3 & 0.76 & 0.30 & 0.20 & 0.77 \\
  \hline
  \rowcolor{LightCyan}
  & Mean & 0.77 & \textbf{0.27} & 0.19 & 0.79 \\
  \hline
  \end{tabular}
  \label{tab:blind_test}%
\end{table}

\section{Results and Discussions}
\label{sec:results}



We perform qualitative analysis and quantitative evaluations on generated tumor crops and discuss the results of blind test by radiologists. The results are summarized below:\smallskip

\begin{table*}[h]
  \centering
  \caption{Mutation detection using only synthetic data generated using SAGE, PGGAN, WGAN-GP and DC-GAN. Data is generated using the training set of shaded row and tested on the corresponding real test set for all 10-folds. The results reported are mean (standard deviation) over 10 folds. Row: cyan is baseline trained on real data. Metrics are - ACC: Accuracy, SEN: Sensitivity, SPEC: Specificity and DIC: Dice Score for IL: Image Level and PL: Patient Level analyses.}

  \begin{tabular}{c|c|ccccc}
  \bfseries Input Type & \bfseries Input source & \bfseries ACC (PL) & \bfseries ACC (IL) & \bfseries SEN (IL) & \bfseries SPEC (IL) & \bfseries DIC (IL)\\ \hline \hline
  \rowcolor{LightCyan}
  & Real Data & 0.85 (0.12) & 0.80 (0.09) & 0.69 (0.09) & 0.84 (0.06) & 0.70 (0.07) \\
  \textit{Tumor} & \textbf{SAGE (ours)} & \textbf{0.85 (0.10)} & \textbf{0.81 (0.07)} & \textbf{0.72 (0.07)} & \textbf{0.87 (0.06)} & \textbf{0.72 (0.08)} \\
  \textit{Crops} & PGGAN & 0.78 (0.15) & 0.72 (0.12) & 0.68 (0.14) & 0.76 (0.10) & 0.72 (0.11) \\
  & WGAN-GP & 0.78 (0.08) & 0.76 (0.09) & 0.70 (0.12) & 0.84 (0.06) & 0.72 (0.09) \\
  &  DC-GAN & 0.74 (0.09) & 0.71 (0.10) & 0.68 (0.12) & 0.82 (0.08) & 0.68 (0.10) \\
  \hline
  \rowcolor{LightCyan}
  & Real Data & 0.92 (0.08) & 0.90 (0.06) & 0.85 (0.07) & 0.95 (0.05) & 0.86 (0.08) \\
  \textit{Whole} & \textbf{SAGE (ours)} & \textbf{0.92 (0.09)} & \textbf{0.89 (0.07)} & \textbf{0.84 (0.08)} & \textbf{0.94 (0.04)} & \textbf{0.86 (0.08)} \\
  \textit{Images} & PGGAN & 0.90 (0.09) & 0.86 (0.09) & 0.80 (0.12) & 0.90 (0.06) & 0.82 (0.09)\\
  & WGAN-GP & 0.85 (0.06) & 0.80 (0.07) & 0.76 (0.09) & 0.88 (0.05) & 0.75 (0.08) \\
  &  DC-GAN & 0.80 (0.08) & 0.78 (0.10) & 0.74 (0.12) & 0.84 (0.06) & 0.77 (0.12) \\
  \hline
  \end{tabular}
  \label{tab:fake-train}%
\end{table*}

\textit{1. Qualitative Evaluation of Generated Images:} We evaluate the ability of SAGE to learn the data distribution accurately. To show that SAGE can learn the features compared to other GANs, the tumor crops and whole slide images generated by SAGE, PGGAN~\cite{han2019learning}, DC-GAN~\cite{radford2015unsupervised} and WGAN-GP~\cite{gulrajani2017improved} are shown in Figure~\ref{fig:sage-res}. We use WGAN-GP over standard WGAN because the gradient penalty has been shown to provide superior results \cite{gulrajani2017improved}. SAGE generates attributes sequentially, giving us complete control over how and what kind of images we generate. It can be seen from Figure~\ref{fig:sage-res} that we are able to recreate specific tumor attributes. For comparison, we cherry-picked the SAGE, PGGAN, WGAN-GP and DC-GAN images that looked visually similar to the corresponding real samples. It can be seen that SAGE is able to recreate the detailed attributes more faithfully. 

Our goal, however, is not limited to recreating features exactly. We also want to generate good quality images and a diverse set of synthetic images that learn from the real distribution. Figure~\ref{fig:diversity-res} shows the diversity of images generated using each of the GANs. It can be seen that DC-GAN suffers from mode-collapse due to lack of data. WGAN-GP is relatively better at handling it, however, given the limited data available, it cannot combat mode-collapse and generate detailed attributes at the same time. Therefore, we can see that while WGAN-GP generates diverse images compared to DC-GAN, it is unable to achieve the same quality of tumors as SAGE. PGGAN evidently performs better than WGAN-GP and DC-GAN but it can be seen that it generates some artifacts in tumor generation. But while PGGAN generates good quality, it is unable to account for diversity. SAGE not only generates high quality images, but also generates more diverse set of synthetic images. Additionally, Figure~\ref{fig:zoomed-res}, zooms in on some examples of generated images where PGGAN, WGAN-GP and DC-GAN fail to generate an acceptable detail and quality of tumor image.

\textit{2.  Quantitative Evaluation:} As a quantitative measure, we calculated the Inception Score of generated images. For two classes, the score can be between 1 and 2. The closer the score is to 2, the better our GAN performs - in terms of realness and diversity. We calculated Inception score for a total of 1000 images (500 each of control and mutated). We calculated SSIM over the same set. We used same settings to evaluate Inception score and SSIM for comparison between SAGE and other GANs. The results are shown in Table~\ref{tab:quality_metrics}. We note that SAGE gives an Inception Score value of 1.71 which shows that it is effective in learning both quality and diversity of data. The SSIM value above 0.5 indicates that visual properties of generated images are similar to real images but since the SSIM is not too close to 1, we can infer that there is a certain level of diversity associated with data.\medskip

\textit{3. Evaluation by Radiologists:} We also report results from a blind test conducted with our radiologists (Table~\ref{tab:blind_test}) to demonstrate that SAGE can generate realistic looking good quality tumors. Referring to Table~\ref{tab:blind_test}, we note a FPR of 35$\%$ and 27$\%$ for control and mutated classes, respectively. We also observe a FNR of 33$\%$ and 19$\%$, respectively. This indicates that SAGE is also able to generate realistic new tumors, i.e., learning how a new tumor would realistically grow. For the tumors that the radiologists could detect as being synthetically generated, we obtained feedback from the radiologists about what features look artificial. There was an agreement among radiologists that an image was computer generated because one of two things: (a) a specific artifact which seemed to be present on some of the images. Many of them had sharp lines or a broken appearance, discontinuity. Those images were suspected as computer generated. They also classified ``smudgy" appearance as computer generated or looking fake, and (b) the brain cortex seemed discontinuous or abnormal on some of the images, which was labeled as computer generated. Further, they did not find anything about the tumors themselves that was different between real and generated images.\medskip

\textit{4. Learning Discriminatory Features:} This test is to demonstrate that SAGE can consistently learn unapparent yet discriminative visual properties of tumors between presence and absence of mutation. Table ~\ref{tab:fake-train} (cyan) reports the evidence of existence of discriminative features between the two classes (consistent high accuracy in mutation detection). Once we establish that mutation can be detected using the classifier, our objective is to evaluate how well SAGE is able to learn from a limited sample space. If synthetic images learn features accurately, then they should be equally effective in detecting mutation. Thus, we replace the training set with synthetic images generated using SAGE. We get similar performance in mutation detection using the synthetic training set as we get using real set. For the experiments in Table~\ref{tab:fake-train}, the dataset was divided patient-wise via a 80-20 split for 10 folds. We train ResNet18 for classification and report mean and standard deviation. We describe two types of results: patient-level (PL) classification and image-level (IL) classification. IL results are the results by considering each slice independently. PL results are obtained by computing the weighted mean of all images of a patient. The rest of Table ~\ref{tab:fake-train} reports the results obtained using a similar training protocol, but trained on various training sets. The three training sets are: generated images using PGGAN, WGAN-GP, DC-GAN and SAGE, respectively. The generated images for 10-folds are learnt from the training sets of the corresponding 10-fold real data and evaluated on corresponding test set. We evaluate whether the generative model can consistently and accurately capture the required distribution to distinguish between control and mutated groups (via a 10-fold cross-validation). The test set for every corresponding fold was kept constant across all the experiments (SAGE, PGGAN, WGAN-GP and DC-GAN) for fair comparison with baseline (cyan, real data). For training we used pre-trained ResNet18 using Adam optimizer, learning rate of 0.0001 with step learning rate scheduler and binary cross-entropy loss. The visual discriminatory features between classes are unapparent and yet classification using SAGE images consistently performs at par with real data. Due to this, we observe that SAGE is able to learn and recreate subtle indicators of tumors that discern the status of mutation.  

\section{CONCLUSIONS}
\label{sec:conclusion}
Medical image generation is a challenging task due to the restrictive nature of data, diversity in individual cases and subsequently complex feature space. In this work, we have shown that synthetic images generated by SAGE match very closely with real data, thereby indicating that SAGE is able to accurately capture data distribution from a limited dataset. Comparing with PGGAN, WGAN-GP and DC-GAN, we can see that SAGE has superior performance in terms of capturing diversity and detail, with an Inception score of 1.71. Upon closer inspection of Figure~\ref{fig:zoomed-res}, we can see that where other GANs become blurry or lose features, SAGE generates detailed tumors. Furthermore, due to the sequential generation of features, we have complete control over the tumors we generate. Therefore, the situations where other GANs overfit and generate the same tumors repeatedly can be avoided in SAGE. The classification tasks indicates that the classifier trained on SAGE images performs just as well as real images, indicating the discriminatory features correlating to presence or absence of mutation are learnt accurately. The quantitative and qualitative evaluation showed that SAGE is an effective approach for generating synthetic data using limited datasets. In addition to these analyses, the visual quality and realism of the images was tested via the blind test. From the blind test it can be concluded that SAGE generates realistic images and not just recreate images from random re-sampling of features. We notice that SAGE has an improvement of 7-12$\%$ as compared to other methods in learning features and a consistently high performance. Summarizing these results, we note that SAGE learns the features and statistical model accurately. We can, therefore, conclude that SAGE can potentially be applied to tackle many disease and anomaly detection problems with limited and/or fragmented data.\smallskip


\noindent \textbf{Acknowledgements.} This research is supported in part by Bourns Endowment funds.

\bibliographystyle{IEEEbib}
\bibliography{strings, refs}





\end{document}